**Monitoring overall survival in pivotal trials in indolent cancers**

**Authors:** Thomas R Fleming*[1], Lisa V Hampson*†[2], Bharani Bharani-Dharan[3], Frank Bretz[2], Arunava Chakravartty[3], Thibaud Coroller[3], Evanthia Koukouli[4], Janet Wittes[5], Nigel Yateman[2], Emmanuel Zuber[2]

* Joint first authorship
† Corresponding author

**Affiliations:**

1. Department of Biostatistics, University of Washington, Seattle, WA, US
2. Novartis Pharma AG, Basel, Switzerland
3. Novartis Pharmaceuticals Corporation, East Hanover, NJ, US
4. Novartis Pharmaceuticals UK, London, UK
5. Wittes LLC, Washington, DC, US

**Corresponding author:** Lisa V Hampson, Novartis Pharma AG, Postfach 4002, Basel, Switzerland. E-mail: lisa.hampson@novartis.com

**Abstract:** Indolent cancers are characterized by long overall survival (OS) times. Therefore, powering a clinical trial to provide definitive assessment of the effects of an experimental intervention on OS in a reasonable timeframe is generally infeasible. Instead, the primary outcome in many pivotal trials is an intermediate clinical response such as progression-free survival. In several recently reported pivotal trials of interventions for indolent cancers that yielded promising results on an intermediate outcome, however, more mature data or post-approval trials showed concerning OS trends. These problematic results have prompted a keen interest in quantitative approaches for monitoring OS that can support regulatory decision-making related to the risk of an unacceptably large detrimental effect on OS. The US Food and Drug Administration, the American Association for Cancer Research, and the American Statistical Association recently organized a one-day multi-stakeholder workshop entitled 'Overall Survival in Oncology Clinical Trials'. In this paper, we propose OS monitoring guidelines tailored for the setting of indolent cancers. Our pragmatic approach is modeled, in part, on the monitoring guidelines the FDA has used in cardiovascular safety trials conducted in Type 2 Diabetes Mellitus. We illustrate proposals through application to several examples informed by actual case studies.

1. <u>Introduction</u>

Indolent or chronic cancers are characterized by long overall survival (OS) time, with median survival exceeding 5-10 years. Examples of indolent cancers include follicular lymphoma, multiple myeloma, and early-stage breast cancer. In indolent cancer, the effects of an experimental intervention on OS could meaningfully influence its benefit-to-risk profile. However, clinical trials in indolent cancer are often inadequately powered to assess effects on OS, due to the limited number of deaths observed during the typical trial duration. Instead, these trials are often designed to support accelerated or full approval decisions by providing reliable evidence about treatment effects on surrogate or intermediate clinical outcomes such as progression-free survival (PFS), disease-free survival (DFS), or objective response rate (ORR). Some of these studies use outcomes that directly measure how a patient feels or functions, such as patient reported outcomes (PROs). Similar considerations also apply to clinical trials performed in early-stage cancers in a curative setting.

Experiences in recent clinical trials conducted in indolent cancers have increased the recognition that some experimental interventions may unfavorably affect OS, despite having favorable effects on the intermediate clinical outcome used to support approval (Merino et al, 2023). In several recently reported studies that had shown reliable evidence of beneficial effects on the trial's primary outcome at the time of the trial's Primary Analysis, the OS data from the same trial (Kumar et al., 2020) or data from subsequent post-approval studies have shown a neutral effect of the intervention on OS or have even been suggestive of harm. A classic example is PI3K-inhibitors, discussed at the April 21$^{st}$, 2022, US Food and Drug Administration (FDA) Oncologic Drugs Advisory Committee (ODAC) meeting, where although several pivotal trials had yielded strongly statistically significant beneficial effects on intermediate clinical outcomes, nonetheless effects on OS were unfavorable (FDA ODAC, 2022; see also Table 1). Such examples have prompted widespread discussion of how best to monitor accumulating OS data for evidence of harm in pivotal clinical trials in indolent cancers where the overarching aim is to "rule out shortening of survival due to toxicity or interference with receiving effective standard / salvage therapies" (Mesa, 2023). See, for instance, materials from the recent joint workshop of the FDA, the American Association for Cancer Research (AACR), and the American Statistical Association (ASA) (FDA-AACR-ASA, 2023).

[Insert Table 1 here]

One important challenge to monitoring OS in trials of indolent cancers is the sparsity of information and the need to make decisions under high levels of uncertainty. Similar issues arise in the rare disease setting, where performing a well powered trial in a reasonable timeframe is often infeasible. Alternative approaches proposed for the design of clinical trials in small populations include frequentist designs with modest power and a higher than conventional false positive error rate (Parmar, Sydes & Morris, 2016; Section 4.3.3 of ICH E11A, 2022). Relaxing the false positive error rate control might be more acceptable if other data is supportive of a benefit on the clinical outcome (O'Connor & Hemmings, 2014). This argument is particularly relevant for our OS monitoring application, where statistically significant results on PFS will likely increase confidence that the experimental intervention has some beneficial, or at least neutral, effect on OS.

Because disease progression is on a causal pathway for OS and individual patient responses on PFS and OS are correlated, there has been considerable interest in whether it would it be possible to increase the precision of our inferences on OS by leveraging data on PFS. While these approaches enable increased precision, a relevant consideration regarding their reliability is that a 'correlate does not a surrogate make' (Fleming & DeMets, 1996; DeMets, Psaty & Fleming, 2020). Several authors have sought to increase the power of a single-stage or group sequential test of the effect of a novel intervention on a long-term clinical efficacy outcome by using the patient-level correlation of baseline prognostic factors and/or short-term outcomes with the long-term response of interest to increase the precision of mid-study estimates of the long-term treatment effect (Galbraith & Marschner, 2003; Marschner & Becker, 2001; Hampson & Jennison, 2013; Van Lancker et al, 2020; FDA, 2023). Importantly, these approaches focus on efficacy testing and presume that performing a highly powered, definitive, test of the long-term effect (albeit, with support from the short-term data) is feasible. These assumptions will be violated in the indolent cancer setting because of the limited number of deaths expected, and since the focus is on monitoring OS as a safety endpoint. In what follows, therefore, we propose OS monitoring guidelines that differ in essential ways from group sequential tests, including handling multiplicity differently. A precedent for this appears in quantitative approaches to signal detection in safety, where the false positive error rate is typically not protected across analyses performed at different timepoints (Zhu et al, 2016).

In the setting of pivotal oncology trials with PFS as the primary outcome, Shan (2023) adapts an approach originally proposed for single-arm trials (Shan, 2021) to develop a guideline for evaluating OS at a single timepoint, that is, the time of the decisive PFS analysis. The first step in designing these guidelines is to specify the power and false positive error rate, then derive the decision thresholds and required number of deaths in light of these. An important challenge is that reaching this number may not be possible. The actual number of deaths occurring depends on a design powered to test the primary outcome, PFS. Having OS monitoring guidelines based on a separate sample size calculation for OS may therefore be impractical to implement. How these guidelines connect to precedents for regulatory decision making in other disease areas is also unclear.

In this paper, we formulate monitoring procedures for OS that are intended not to provide 'rules' but rather, informative guidelines and corresponding 'thresholds for positivity', for regulatory authorities and Data Monitoring Committees when judging whether emerging evidence about effects on OS is concerning or not. An important decision point occurs at the time of the trial's primary outcome analysis, which is intended to assess trial success, potentially leading to submission and thereafter possibly regulatory approval. We refer to this analysis as the 'Primary Analysis' and its calendar timing as $t_{PA}$. If the primary outcome is tested according to a group sequential design, the time of the Primary Analysis typically would be the calendar time when the group sequential boundary is crossed and a DMC recommendation for termination of testing for the primary outcome is accepted. A second important decision milestone is the 'Final Analysis', which occurs at calendar time, $t_{FA}$, when the final analysis of OS is conducted (which may occur post-marketing if approval is based on available data at $t_{PA}$). Figure 1 provides a schematic diagram illustrating the sequencing of important decision milestones in our proposed monitoring guidelines.

[Insert Figure 1 here]

We emphasize that our proposed OS monitoring guidelines differ fundamentally from classical group sequential boundaries (O'Brien & Fleming, 1979; Jennison & Turnbull, 2000; Proschan, Lan & Wittes, 2006; Wassmer & Brannath, 2016). Such boundaries are typically implemented to provide a definitive test of superiority of efficacy; they are designed with high power to achieve statistical significance under a clinically relevant effect size. In particular, the group sequential boundary provides guidance about whether data at interim analyses are sufficiently extreme to enable early termination of the trial, either because these data definitively establish or rule out that the experimental intervention has a beneficial effect.

In contrast, we consider OS monitoring guidelines for settings where the trial is unlikely to provide definitive evidence about whether the experimental intervention provides a clinically meaningful beneficial effect on OS at time $t_{FA}$, even if this occurs long after the Primary Analysis. Instead, a 'favorable' result for OS could be successfully ruling out a rigorously defined minimum detrimental effect on OS using evidentiary thresholds that are more lenient than the requirement for statistical significance adjusted for multiple testing over time that is conventional for confirming efficacy. The minimum unacceptable detrimental effect would be set considering the beneficial effect that must be observed on the trial's primary outcome to achieve statistical significance. At the time of the Primary Analysis, the proposed OS monitoring guidelines are intended to guide judgments about whether results are adequately favorable rather than justifying early trial termination with a definitive conclusion.

In this paper, we propose and illustrate monitoring guidelines for OS that are tailored to clinical trials conducted in indolent cancers, modeled in part on the monitoring guidelines used by FDA in settings of cardiovascular safety trials conducted in Type 2 Diabetes Mellitus (T2DM) or in severe obesity. While we concentrate on settings where obtaining both a timely and reliable assessment of the effect on OS would be infeasible, our proposed guidelines would also be relevant in scenarios where only limited insights about effects on OS would be available at the time of the Primary Analysis, and yet OS could be tested for superiority as a pre-specified secondary outcome maybe 1.5 to 5 years later, at the Final Analysis of OS, albeit with modest power. In such situations, our proposed methods could be particularly useful in determining whether unacceptably large detrimental effects on OS could be ruled out, especially at the Primary Analysis of the primary outcome. Table 2 provides a glossary of the terminology and notation used throughout this paper.

2. Motivating examples

We motivate the subsequent work in this paper through several examples that highlight the practical challenges associated with monitoring OS; we will return to them in later sections to illustrate our proposed OS monitoring guidelines. For reasons of confidentiality, we have anonymized some examples. Throughout we assume that the hazard ratio (HR; test/control) can appropriately summarize the magnitude of benefits and harms of the experimental intervention compared to control.

**Example 1:** POLARIX was a Phase 3 randomized controlled trial (RCT) allocating 879 adults with untreated large B-cell lymphoma (LBCL) 1:1 between polatuzumab vedotin + R-CHP and placebo + R-CHOP (i.e., the experimental intervention substituted one element of R-CHOP, vincristine, with polatuzumab vedotin). The primary outcome was PFS with OS a key secondary outcome (i.e., a secondary outcome reflected in the trial's hierarchical testing strategy to control the familywise false positive rate); if OS could be formally tested, the Final OS Analysis was expected to be based on 178 deaths. The protocol specified that the proposed group sequential design for monitoring OS with looks at 134 and 178 deaths would have power 0.52 to detect a HR of 0.73 at the Final OS Analysis. The sponsor subsequently calculated that 631 deaths would be needed for power 0.8 to declare superiority on OS if the HR for OS was 0.8 (see pp.96-97 of combined briefing document; FDA ODAC, 2023).

POLARIX provided statistically significant evidence of a benefit on PFS (HR estimate = 0.73; 95% CI 0.57 to 0.95), but with "modest" benefit on the probability of PFS at 2 years (6.5% difference) (see pp.15 and pp.42 of combined briefing document; FDA ODAC, 2023). Only 131 deaths occurred, instead of the planned 178, at the Final OS Analysis. While the observed HR for OS among all randomized patients was slightly below 1, (HR estimate= 0.94; 95% CI 0.67 to 1.33), substantial uncertainty remained about the effect of the experimental intervention on OS. Consequently, the FDA requested the sponsor provide additional analyses quantifying the potential for observing evidence of a detrimental effect on OS in the overall population with 2 or 5 years of additional follow-up. The conditional probability that the point estimate of the OS HR would exceed 1 after 5 years additional follow-up was calculated to be 0.05 if the true OS HR were 0.8, and 0.6 if it were 1.1, (Briefing materials for ODAC meeting, March 2023), so that even this additional follow-up might not be sufficient to rule out OS harm reliably. At the FDA ODAC meeting in March 2023, members of the Committee judged that more than 10 years additional follow-up would likely be needed to rule out harm reliably, which is infeasible in this front-line setting. In April 2023, polatuzumab vedotin + R-CHP received full approval.

This example highlights the challenge of accruing definitive levels of evidence on OS in a timely manner even in the frontline setting of an aggressive cancer type. Similar challenges are encountered in the indolent cancer setting, as illustrated in the next example.

**Example 2:** Consider an indolent cancer setting with a confirmatory RCT designed to recruit 108 adult patients in 27 months and randomize patients 1:1 between an experimental intervention and control. The trial must accrue at least 66 PFS events to have power 0.8 to declare superiority of the experimental intervention versus control when the true PFS HR is 0.5; statistical significance on the primary outcome will be reached at the final PFS analysis if the PFS HR estimate ≤ 0.62. The team assesses a 30% reduction in mortality risk to be a plausible benefit of the intervention on OS.

Under the planned design, the time of the first OS analysis would coincide with the decisive PFS analysis after 55 months, when 22 deaths would be expected (timings and event numbers calculated under the alternative hypothesis assuming: exponential times-to-event, and theoretical loss-to-follow-up rates for PFS and OS events of 10% and 1% per year, respectively). It would be considered reasonable, in the trial scientific and drug development context, to schedule the final OS analysis after 87 months of patient follow-up; at that time 34 deaths would be expected. As a point of reference, defining an OS HR ≥ 1.333 as a clinically relevant detrimental effect in light of the disease setting and available external data, a total of 102 deaths would be required to rule out $H_0$: OS HR = 1.333 at one-sided significance level 0.025 with power 0.9 assuming the OS HR = 0.7. This example highlights that RCTs in indolent cancers, with a sample size tailored to the primary surrogate or intermediate clinical outcome, will likely be severely underpowered to rule out clinically relevant detrimental OS effects if the final OS analysis is scheduled at a calendar time considered reasonable in the context of the overall development program.

**Example 3:** The BELLINI trial (Kumar et al., 2020), a pivotal RCT, randomized 291 patients in 2:1 ratio between venetoclax (a targeted therapy) or placebo as an add-on to bortezomib and dexamethasone in patients with relapsed/refractory multiple myeloma who had received 1 to 3 previous lines of treatment. The primary outcome was PFS, with OS a key secondary outcome. The analysis of PFS was based on the data snapshot taken on November 26$^{th}$, 2018. At this cutoff, while promising results were observed for PFS (HR estimate 0.63; 95% CI 0.44 to 0.90) in the all-comer population, evidence suggesting harm was observed for OS (HR estimate 2.03, 95% CI 1.04 to 3.95) in this full group. Divergent trends were observed in a biologically plausible subgroup of patients with t(11;14) gene expression (OS HR estimate 0.34, 95% CI 0.03 to 3.84), although this subgroup analysis was not pre-specified and had only 35 patients of whom three had died. Results from such subgroup analyses have limited reliability, because of multiple testing, and usually should be viewed as being hypothesis generating (EMA, 2019; Alosh et al., 2017). The recent AACR-FDA-ASA workshop on OS in oncology trials cited the BELLINI trial, highlighting the importance of carefully considering OS, overall and in important subgroups, and the challenge of sparse information when interpreting potentially divergent OS trends across populations (Gormley, 2023).

3. <u>Formulating a pre-specified OS monitoring guideline</u>

This section focuses on prospectively designing a 'formal' OS safety monitoring guideline to address the risk of meaningfully unfavorable effects, where 'formal' refers to prespecifying decision criteria. Our proposals are inspired, in part, by monitoring guidelines suggested by the FDA Endocrinologic & Metabolic Drugs Advisory Committee (EMDAC) in 2008 in the setting of cardiovascular safety trials conducted in patients with T2DM or severe obesity.

*3.1.   Guidelines for monitoring major morbidity/mortality effects in the T2DM setting*

The challenges in monitoring irreversible morbidity/mortality effects in the T2DM setting are similar in some respects to those in monitoring OS effects in indolent cancer settings. While regulatory approvals in T2DM prior to 2008 had been based primarily on evidence about beneficial effects on the biomarker $H_bA_{1c}$, emerging evidence indicated that some T2DM agents might have unfavorable effects on macrovascular complications, such as cardiovascular (CV) death, stroke, and myocardial infarction (MI). In its July 2008 meeting, the FDA EMDAC recommended conducting randomized trials with the primary efficacy MACE composite outcome, (Major adverse cardiovascular event, i.e., CV death, stroke, MI). The recommendation was to design those trials to rule out a MACE HR of 1.3 with a one-sided false positive error rate 0.025, and with power 0.9 at the Final Analysis if the true MACE HR were 1.0. This required the clinical trial to have 611 MACE events assuming 1:1 randomization between the experimental intervention and control. EMDAC also recommended basing marketing approval on efficacy and safety evidence that established persuasively favorable effects on $H_bA_{1c}$. At this time, the sponsor was also required to report a current analysis of MACE that was based on the occurrence of at least 122 MACE events, (i.e., essentially the first 20% of trial information on this outcome). An analysis based on those 122 events would have power 0.9, when the true MACE HR is 1.0, to rule out a MACE HR of 1.8 with one-sided false positive error rate 0.025. The FDA endorsed these EMDAC recommendations and subsequently issued a guidance that drew on these (FDA Center for Drug Evaluation, 2008). More broadly, still assuming 1:1 randomization, if there were $L_{PA}$ MACE events at the Primary Analysis conducted to support marketing approval, then the HR 'margin of detriment' that would need to be ruled out would be exp $\{(1.96+1.282) \sqrt{4/L_{PA}}\}$. In turn, the threshold for positively ruling out that margin would be an estimated HR of exp $\{1.282 \sqrt{4/L_{PA}}\}$.

Important features of this monitoring approach in the T2DM setting included the pre-specification of the required levels of precision for the estimated MACE HR, both at the time of marketing approval as well as at the Final Analysis, and pre-specification of the margins to be ruled out at both the Final Analysis and the Primary Analysis.

This approach acknowledges the substantial uncertainty inherent in the limited MACE data available at the Primary Analysis by aiming for a more modest objective of reliably ruling out a larger detrimental effect at this stage. Alternatively, this more modest objective could be reframed as needing to rule out the MACE HR of 1.3, considered to be a clinically relevant detrimental effect in this setting, using a more lenient (i.e., higher) false positive error rate. Specifically, assume 1:1 randomization and suppose the Primary Analysis will be performed at $L_{PA}$ MACE events and define the smallest unacceptable detrimental HR (denoted by $\delta_{null}$) as 1.3 in this MACE setting. Then the criterion of ruling out a MACE HR of $\exp\{(1.96+1.282)\sqrt{4/L_{PA}}\}$ at one-sided false positive error rate 0.025 can be reformulated as the criterion of ruling out $\delta_{null}$ at one-sided false positive error rate $\gamma_{PA} = 1 - \Phi\left(\frac{\sqrt{L_{PA}}\log(\delta_{null})}{2} - 1.282\right)$. In this setting where $L_{PA}$ = 122, then 1.26 is the estimated MACE HR that is the threshold for ruling out the HR = 1.8 at a one-sided false positive error rate 0.025. It is also the threshold for ruling out the HR =1.3, (i.e., $\delta_{null}$), at a one-sided false positive error rate $\gamma_{PA} = 0.43$. Suppose, however, that the Primary Analysis had been conducted at 50% (instead of 20%) information, i.e., when $L_{PA}$ = 306 MACE events. Then the margin to be ruled out with one-sided 0.025 false positive error rate would be a HR = 1.45; the threshold for positively ruling this out would be an estimated HR 1.158. In turn, this threshold would rule out the clinically relevant detrimental HR = 1.3, (i.e., $\delta_{null}$), at a considerably lower one-sided false positive error rate $\gamma_{PA} = 0.15$, thereby directly reflecting the reduced uncertainty achieved with the additional information.

The advantage of phrasing the criterion at every analysis in relation to the same level of detriment, $\delta_{null}$, is that it makes explicit the proposed trade-off between the risks of false positive and false negative decisions at the Primary Analysis. This is particularly relevant for indolent cancers where considerations of feasibility mean trials will rarely be fully powered to rule out, at a conventional level of significance, the clinically specified margin representing the smallest unacceptable detrimental HR, meaning such trade-offs will need to be carefully discussed and aligned upon. We therefore favor this framing when formulating OS monitoring guidelines tailored to indolent cancer.

3.2. *Formulating OS monitoring guidelines tailored to the setting of indolent cancers*

Two parameters will play a central role in the design of OS monitoring guidelines: 1) $\delta_{null}$, representing the smallest unacceptable detrimental OS HR; and 2) $\delta_{alt}$, a *plausible* OS HR consistent with (perhaps incremental) benefit that is reasonably expected from the experimental intervention. Pre-specification of both parameters should be based on clinical considerations such as the disease stage (advanced disease vs early-stage disease with potentially curative treatment), expected median OS on control, and the strength of the observed treatment effect on the primary outcome, such as PFS, needed to achieve statistical significance (Mesa, 2023). Discussions may also be informed by relevant data on drugs in the same class or with a related mechanism of action. Both parameters, however, should be set independently of the reasonable number of deaths, as the parameters should represent realistic quantities of interest for patients, physicians, and regulators, specific to the disease setting and the treatment, and independent of study design. Once $\delta_{null}$ and $\delta_{alt}$ are in place, the sponsor can define the timing of the Final Analysis. Suppose a definitive test ruling out $H_0$: OS HR $\geq \delta_{null}$ with one-sided false positive error rate 0.025 and with power 0.9 under $\delta_{alt}$ would require L* deaths. As illustrated in Section 2, these evidentiary levels, while conventional for efficacy evaluations, are not feasibly achieved for an assessment of potential unfavorable effects on OS in indolent cancer settings. If this is the case, the sponsor must justify the number of deaths, $L_{FA}$, that *would* be feasible to occur in a reasonable time ($t_{FA}$) given that the trial sample size is usually tailored to the primary outcome, such as PFS. As in the T2DM setting, we assume the trial will continue (possibly post-approval) to time $t_{FA}$ and reach the OS event target $L_{FA}$. Careful specification of $L_{FA}$ is critical since it will determine the strength of evidence with which to rule out $δ_{null}$ at the Final Analysis.

The 'threshold for positivity' at any given analysis is the value below which the observed HR must be to provide sufficient reassurance that the effect on OS does not reach the selected unacceptable level of detriment (the margin $δ_{null}$). The threshold for positivity will always be less than the selected level, $δ_{null}$. Any observed HR value below the threshold will be regarded as having 'met' the required criterion for acceptable HR for OS at that analysis. A false positive error arises at an analysis when the HR truly is $δ_{null}$ and yet the data meet the threshold for positivity, while a false negative error arises when the HR truly is $δ_{alt}$ and yet the data fail to meet the threshold for positivity.

The proposed monitoring guidelines are intended to flag potentially concerning evidence on long-term safety rather than motivate early termination of the trial. Therefore, an observed OS HR exceeding, (i.e., not meeting), the positivity threshold at the Primary Analysis would prompt careful evaluation of the data in the context of the totality of the evidence available; this review would then inform the action to be taken. In extreme circumstances, one might consider terminating the trial; the more likely course of action would be to continue the study to the Final Analysis to gather more information on OS. On the other hand, if the observed OS HR lies below, (i.e., does meet), the positivity threshold at the Primary Analysis and the novel intervention is submitted for and granted regulatory approval, the trial generally should continue to the Final Analysis to accrue more information on OS.

Assume, for ease of presentation, that two OS analyses are preplanned, one at the Primary Analysis and the other at the Final Analysis. Then, formulation of the monitoring guideline for OS would be based on the steps shown below. See Appendix A for derivations of OS HR thresholds for positivity.

A. Select the one-sided false positive error rate ($\gamma_{FA}$) that can be tolerated for testing $H_0$: OS HR $\geq \delta_{null}$ at the Final Analysis. Denote by $1 - \beta_{FA}$ the attained power, when the true HR equals $\delta_{alt}$, to rule out $\delta_{null}$ at false positive error rate $\gamma_{FA}$ based on $L_{FA}$ deaths. Note that $\gamma_{FA}$ should not be confused with the conventional significance level $\alpha$ (i.e., 0.025): Instead, $\gamma_{FA}$ should be set in consultation with regulatory authorities and with consideration of the related false negative rate $\beta_{FA}$ given that feasibility will often fix the number of deaths $L_{FA}$.

B. Let 'positivity' at the Final Analysis refer to ruling out an OS detriment $\delta_{null}$ at the agreed upon false positive error rate ($\gamma_{FA}$). Then the criterion for positivity at the Final Analysis can be phrased as 'the upper bound of the $(1 - 2\gamma_{FA})100\%$ two-sided CI for the OS HR should be less than $\delta_{null}$'. Equivalently, this can be expressed as 'the point estimate of the OS HR should be less than $\delta_{null} \exp\left\{\frac{2\Phi^{-1}(\gamma_{FA})}{\sqrt{L_{FA}}}\right\}$', termed the 'threshold for positivity', at the Final Analysis.

C. At the Primary Analysis, when the decisive analysis of the primary outcome is performed, suppose we expect $L_{PA}$ deaths to have occurred. If the trial's primary success criteria are met at this milestone, the evidence of a beneficial effect on the primary outcome should be clinically and statistically compelling; typically, $t_{PA}$ will be driven by these considerations and this, in turn, will determine $L_{PA}$.

- Calculate the threshold for positivity, where again 'positivity' at this timepoint corresponds to ruling out $\delta_{null}$ at a relaxed evidentiary level. We calculate the threshold for positivity such that the probability the OS HR estimate lies below this threshold when the true HR equals $\delta_{alt}$ is 1- $\beta_{PA}$. A typical choice for $\beta_{PA}$ may be 0.1 or 0.2 but may be higher if $L_{PA}$ is very low. In the examples we have considered, setting $\beta_{PA} \leq \beta_{FA}$ ensures positivity thresholds at successive analyses form a decreasing sequence, although this is not necessary. The OS HR threshold for positivity is $\delta_{\text{null}}\exp\left\{\frac{2\Phi^{-1}(\gamma_{PA})}{\sqrt{L_{PA}}}\right\}$, where $\gamma_{PA} = 1 - \Phi\left(\frac{\sqrt{L_{PA}}\log(\delta_{\text{null}}/\delta_{\text{alt}})}{2} - \Phi^{-1}(1 - \beta_{PA})\right)$.

- Use of this threshold for positivity implies willingness to accept a (one-sided) false positive error rate of $\gamma_{PA}$ for the assessment of possible OS detriment at the Primary Analysis. To make this (and thus the level of uncertainty under which we are making decisions) more explicit, we could rephrase the criterion for positivity as 'the upper bound of the $(1 - 2\gamma_{PA})100\%$ two-sided CI for the OS HR should be less than $\delta_{\text{null}}$'.

Note that while we formulate guidelines monitoring the OS HR estimate, at each analysis it may be helpful to complement this summary measure by also presenting estimates of supplementary estimands, such as the difference in survival probabilities or in restricted mean survival times at clinically relevant landmark times.

In the above formulation, we set the positivity threshold at the Primary Analysis to ensure adequate power under $\delta_{\text{alt}}$, usually requiring acceptance of an increased false positive error rate. This is justified in part because the trial will continue and will eventually reach the Final Analysis when we will test $H_0$: OS HR $\geq \delta_{\text{null}}$ controlling the false positive error rate at the agreed, more stringent, level of $\gamma_{FA}$. As a technical aside, expressions for the 'thresholds of positivity' shown above are derived assuming OS log-HR estimators follow their asymptotic normal distributions, approximating the precision by one quarter of the number of observed deaths (pooling across trial arms), which should be adequate when the OS log-HR is close to 0 and under 1:1 randomization (Fleming & Harrington, 1991; Jennison & Turnbull, pp.78 and references therein). See Appendix

A for expressions for the positivity thresholds under k:1 randomization. When very few deaths are expected, simulation may be useful to verify the small sample accuracy of the asymptotic distribution and the impact of deviations on operating characteristics. While estimates of the OS log-HR can be based on log-rank statistics or extracted from fitted Cox proportional hazards models, more efficient estimates may be obtained from joint models of OS and repeated biomarker measures if these are understood to be predictive of a patient's time to death (Burdon, Hampson, Jennison, 2023).

The OS monitoring guideline above has two features beyond the focus of the T2DM monitoring procedure. First, the Primary Analysis provides a focus on power at a *user-specified* alternative OS HR $\delta_{alt}$ (which might differ from the OS HR at which the Final Analysis has power 0.9). This is motivated by the fact that in the indolent cancer setting, when $L_{FA}$ may be far below $L^*$, it is important to avoid powering the Primary Analysis at an alternative HR that may correspond to an implausible OS benefit. Second, this approach references the same margin of harm ($\delta_{null}$) at every OS analysis. By referencing the same margin of harm at the Primary Analysis as at the Final Analysis but relaxing the false positive error rate, the guideline makes explicit the impact of the lower information at $t_{PA}$ on the reliability of decision-making. This presentation can facilitate discussions about the acceptability of the more lenient evidentiary threshold at the Primary Analysis.

Sponsors should evaluate and document the operating characteristics (OCs) of an OS monitoring guideline, ideally at the design stage. Relevant OCs may include, but are not limited to the following:

1. The false positive probability of ruling out $\delta_{null}$ at final OS analysis when in fact the true OS HR equals $\delta_{null}$.

2. The false negative probability of not ruling out $\delta_{null}$ at final OS analysis when in fact the experimental intervention has a beneficial or no effect on OS, for example, under the alternative OS HR of $\delta_{alt}$.

3. The false positive probability the OS HR point estimate lies below the OS HR threshold for positivity at the Primary Analysis when in fact the true OS HR equals $\delta_{null}$.

4. The false negative probability the OS HR point estimate exceeds the threshold for positivity at the Primary Analysis (potentially delaying approval) when the true OS HR equals, for example, $\delta_{alt}$.

4. *Illustrating the OS monitoring guidelines*

This section uses three examples to illustrate the proposed guideline.

*4.1 Revisiting motivating example 1: A setting with larger numbers of deaths, $L_{PA}$ & $L_{FA}$*

We retrospectively designed an OS monitoring guideline for the POLARIX trial. The original design of the POLARIX trial specified the Final Analysis of OS to occur when $L_{FA}$ = 178 deaths. For the purposes of illustration, we set $\delta_{null}$ = 1.3, identical to the margin accepted by FDA as the margin of harm at the Final Analysis of the monitoring guideline in the setting of cardiovascular safety trials in T2DM. In practice, this threshold should be clinically selected on a case-by-case basis and aligned with the relevant health authority. We set $\delta_{alt}$ = 0.80, the plausible alternative OS HR the sponsor considered at the time of the FDA ODAC meeting in March 2023 (see pp.97 of combined briefing document; FDA ODAC, 2023). At the Final Analysis, we set the false positive error rate as $\gamma_{FA}$ = 0.025: with $L_{FA}$ = 178 deaths, we then have probability 0.9 of meeting the positivity threshold when the true OS HR is $\delta_{alt}$. We assume this represents an adequate trade-off between the risks of false positive and false negative errors in this setting. At a Primary Analysis when $L_{PA}$ = 110 or 131 deaths, (i.e., at the 62% or 73% information fraction, respectively), we set $\beta_{PA}$ = 0.10 so that the thresholds for positivity shown in Table 3 would be met with marginal probability 0.90 under $\delta_{alt}$ = 0.80. (To illustrate properties of the monitoring boundary if the Primary Analysis were performed even earlier, i.e., at the 33% or 50% information fraction, Table 3 also shows the proposed OS monitoring guideline at $L_{PA}$ = 60 or 89 deaths.)

This approach has favorable operating characteristics even though fewer than the assumed 178 deaths at the Final Analysis occurred. If the Primary Analysis is performed at 62% information fraction when $L_{PA}$ = 110 deaths, the probability of successfully obtaining an OS HR estimate < 1.02 is 0.9 if the true OS HR is 0.80. That is, the study has probability 0.9 of concluding the intervention does not unacceptably increase mortality. Moreover, the probability is still 0.83 if the true OS HR is 0.85. Importantly, the study also has nearly 0.9 probability of not meeting the threshold for positivity when the true OS HR is 1.3. If the detrimental effect is more marked, say the true OS HR is 1.5, the probability the OS HR < 1.02 is only roughly 0.02.

In turn, if the Primary Analysis occurs at the 74% information fraction when $L_{PA}$ = 131, the probability of successfully obtaining an estimate < 1.00 would be 0.9 if the true OS HR is 0.80 and is less than 0.07 if the true OS HR is 1.3. At the final OS analysis of POLARIX (based on 131 deaths), the observed OS HR estimate was 0.94 (95% CI 0.67 to 1.33); this would have met the proposed threshold for positivity of 1.00.

*Nota bene*: The false positive error rate to rule out the HR $\delta_{null}$ = 1.3 in this illustration where the trial has marginal power 0.9 for positivity at each analysis, is the same at each information time as would be obtained when using the MACE boundary in the T2DM setting. Hence, there is clear precedent for the level of false positivity at each analysis in this illustration.

[Insert Table 3 here]

Readers should be cautious in interpreting our proposed methods for monitoring OS when applied in a setting such the POLARIX trial. By intention, these methods would be applied in a setting where the effects at the Primary Analysis on the 'feels, functions' outcome or an intermediate clinical response, such as PFS, would be statistically robust and clinically compelling. Since such results often also would be suggestive of OS benefits, the actual OS results at times $t_{PA}$ and $t_{FA}$ properly achieve the 'threshold for positivity' in our OS monitoring guideline unless they provide substantive evidence either against benefit or suggestive of harm. In contrast, in the POLARIX trial, where the PFS results at the Primary Analysis were consistent with a modest beneficial effect and were not statistically compelling, (supporting FDA's engagement of ODAC), the 0.94 HR for OS is more concerning since, while it suggests lack of harm, it further suggests that overall efficacy is not meaningfully beneficial.

*4.2 An example with intermediate numbers of deaths, $L_{PA}$ & $L_{FA}$*

While the properties of the proposed approach are favorable when many deaths have been reported at the Primary and Final Analysis, in many indolent cancer settings the number of deaths that would be likely to occur at these analyses could be considerably smaller. Continue to suppose that the margin, $δ_{null}$, is 1.3 and the randomization is 1:1. Assume $δ_{alt}$ = 0.70, and $β_{PA}$ = 0.10. Furthermore, suppose that a one-sided false positive error rate of 0.10 is regarded as acceptable at the Final Analysis. In contrast to the previous example, we now consider more realistic targets in indolent cancer. Set $L_{FA}$ = 70 deaths at the Final Analysis and suppose at the Primary Analysis there are $L_{PA}$, = 28 or 42 deaths, i.e. the 40% or 60% information fraction. See Table 4 for the corresponding details for this scenario, including the proposed OS monitoring guideline.

[Insert Table 4 here]

In this scenario, if the Primary Analysis occurs approximately at the 40% information fraction, then the OS HR threshold for positivity would be 1.14. If, however, the Primary Analysis would be delayed until reaching the approximate 60% information fraction, the threshold for positivity would be 1.04. Further, at the Final Analysis, the threshold for having results sufficiently favorable to rule out the 1.3 margin would be an estimated OR HR = 0.96.

This approach has acceptable statistical properties even when one has only $L_{FA}$ = 70 deaths at the Final Analysis, especially if the Primary Analysis would occur at the 60% information fraction. To be specific, at that analysis with $L_{PA}$ = 42 deaths, the probability of successfully obtaining an estimate < 1.04 would be 0.9 if the true OS HR is 0.70 and still is 0.80 if the true OS HR is 0.80; the one-sided false positive error rate is only 0.23 if the true OS HR is 1.3. In turn, at the Final Analysis with $L_{FA}$ = 70 deaths, the probability of successfully obtaining an estimate < 0.96 would be 0.90 if the true OS HR is 0.7 and still is 0.77 if the true OS HR is 0.80; the false positive rate at the Final Analysis is 0.10 if the true OS HR is 1.3.

In another variation on this example, suppose that the experimental intervention is expected to offer important clinical benefits for patients that OS would not directly capture, such as a milder side effect profile, or replacing or delaying transplant and its associated comorbidities. In this case, suppose the intervention would still offer a clinically meaningful benefit for patients even if the OS HR = 0.95. Thus we assign $\delta_{alt}$ = 0.95. Table 5 provides an OS monitoring guideline for this setting, further relaxing the one-sided false positive error rate at the Final Analysis to $\gamma_{FA}$ = 0.2 to ensure an adequate probability of meeting the positivity threshold when the true OS HR is 0.95. In light of the limited power available, we also relax our target power at the Primary Analysis to $1 - \beta_{PA}$ = 0.75. This example highlights that when little, if any, effect on OS is plausible, it is very challenging for any monitoring guideline to achieve an acceptable trade-off between false positive and false negative error rates. In practice, there will typically be uncertainty about what a plausible clinically relevant OS benefit might be. OS monitoring guidelines corresponding to different values of $\delta_{alt}$ (including values characterizing incremental benefits such as $\delta_{alt}$ = 0.95, or even $\delta_{alt}$ = 1 if appropriate) should be systematically made available at the time of the Primary and Final Analyses for context and to support caution in following the guidelines too definitively.

[Insert Table 5 here]

*4.3 Revisiting motivating example 2: A setting with inherently smaller $L_{PA}$ & $L_{FA}$*

Given the inherent lack of statistical power in this clinical setting, suppose the study stakeholders agree it is sufficient to rule out $\delta_{null}$ = 1.333 with a one-sided false positive error rate 0.20 at the Final Analysis, and we want to meet the positivity threshold at the Primary Analysis with probability 0.9 under $\delta_{alt}$ = 0.7. Here, $\delta_{null}$ is similar to, but slightly larger than, the 1.30 margin used in the T2DM setting: We assume that the magnitude of benefit on the primary outcome of PFS needed for statistical significance, that is a PFS HR estimate < 0.62, represents a substantive benefit for patients, such that a larger value of $\delta_{null}$ is justified in this case, In the notation of Section 3, for Motivating Example 2 we have $L_{PA}$ = 22, $L_{FA}$ = 34, $\gamma_{FA}$ = 0.20 and $\beta_{PA}$ = 0.1. Table 6 shows the OS monitoring guideline applied to this example.

[Insert Table 6 here]

In this scenario, if the Primary Analysis would occur approximately at the 65% information fraction, then the OS HR threshold for positivity would be 1.21. At the Final Analysis, the threshold for having results sufficiently favorable to rule out the 1.333 margin would be an estimated OS HR of 0.999. The probability is 0.15 of failing to meet the positivity threshold at the Final Analysis under an OS HR of 0.7. If the OS HR is 0.85, the probability is 0.32. These false negative error rates at the Final Analysis under what would be considered as promising beneficial OS effects are certainly much higher than would be tolerated in other settings. As we noted in Section 3.2, consideration of the false negative error rate $\beta_{FA}$ should play a central part in the specification of $\gamma_{FA}$. In cases where only small or intermediate numbers of deaths will have occurred by the time of the Final Analysis, power may become the most prominent consideration, so that $\gamma_{FA}$ is instead fixed by the choice of a tolerable false negative error rate $\beta_{FA}$.

5. Discussion

In this paper, we propose novel OS monitoring guidelines to use in indolent cancer settings or other similar contexts such as clinical trials of early-stage cancers in a curative setting. See the monitOS R package (https://cran.r-project.org/web/packages/monitOS/index.html) for R code implementing the approaches of this paper; additional package documentation can be found at https://opensource.nibr.com/monitOS/. This approach is primarily intended to provide pragmatic, yet rigorous, guidance to drug developers and regulatory decision-makers on what would be acceptable evidence on OS, assuming statistically and clinically compelling benefits are demonstrated on the primary outcome, and where interpretation of OS would also be appropriately influenced by other relevant data such as non-fatal safety events. The DMC should be informed about these proposed OS monitoring guidelines, ideally at their DMC Organizational Meeting, even though the principal role of the guidelines would not be to trigger a DMC decision about potential early termination.

We have shaped our proposed OS guidelines using three fundamental ingredients: 1) the clinically determined margin of detriment on OS that is unacceptably high; 2) the benefit on OS that is plausible given the mechanism of action of the experimental intervention; and 3) the quantity of information it is feasible to accrue given the clinical and drug development setting. When considering the first ingredient, $\delta_{null}$, patients may be willing to trade some increase in their long-term risk (represented by an OS detriment) for short-term benefits such as reduced toxicity, improvements in how they feel or function as captured by PROs, and proven benefits on the primary outcome. When discussing such trade-offs, it may be informative to translate $\delta_{null}$, representing an increase in the relative risk of death, to a difference in absolute risk. In the indolent cancer setting, this may reveal that $\delta_{null}$ corresponds to a moderate absolute increase in long-term risk, which should then be weighed against substantial short-term benefits. Patient preference studies, such as discrete choice experiments, (IMI PREFER, 2021, and references therein), could be used to quantify the maximum trade-off acceptable to patients and thus inform specification of $\delta_{null}$. Evidence on short-term benefits should be accounted for when interpreting OS results at the Primary and Final Analysis. Strong evidence of a benefit on PROs could further support the approval of a medicine at the time of the Primary Analysis with an observed OS HR > 1.

The proposed guidelines facilitate transparent discussions between stakeholders focusing on the risks of erroneous decisions and what might be an acceptable trade-off between power and the false positive error rate at each analysis. The proposed guidelines bear important differences from classic group sequential designs. When using group sequential designs, crossing the boundary at a single analysis may be sufficient to justify early trial termination, either by establishing or by ruling out benefit, and hence multiplicity adjustments are made to protect error rates across analyses. In contrast, the proposed OS safety monitoring guidelines would inform judgments about whether results are adequately favorable at both the time of the Primary Analysis when there would be $L_{PA}$ deaths, as well as (typically in a post-marketing setting) at the time of the Final Analysis when there would be $L_{FA}$ deaths. Multiplicity is addressed in a different manner in this setting since we do not attempt to control the false positive or false negative error rates across analyses, where some parallels can be drawn with how multiplicity is typically handled when monitoring unexpected or rare serious safety events in clinical trials (Proschan, Lan & Wittes, 2006).

Clearly the requirement to rule out concerning evidence of an unacceptable detrimental OS effect adds an additional hurdle for both trial success and regulatory approval. While we emphasize that the OS monitoring guidelines should be applied as just that – guidelines and not rules – prespecifying them enables sponsors to support more transparent discussions on uncertainties on the drug effect on OS. Their pre-specification also enables the sponsor to quantify the potential risk of failing on OS and incorporate this into the evaluation of the probability of success, thus more effectively managing stakeholder expectations.

In many, if not most, indolent cancer settings, the trial will be inherently underpowered for assessments of OS, such as in Example 2 where there would be only $L_{FA}$ = 34 expected deaths even at the Final Analysis. In such settings, a significant challenge would be justifying the more lenient evidentiary standard at which to rule out the margin, $δ_{null}$, at the Final Analysis. Acceptance of a larger false positive error rate at this analysis is, in part, necessary to avoid excessive risks for false negative errors in clinical settings where it is inherently difficult to increase $L_{FA}$ to be considerably more than 34 deaths. Relaxing the one-sided false positive error rate to be larger than 2.5%, such as 5% or even 10% to 20%, might be justified if the primary outcome is a clinically important 'feels, functions' measure and a necessary condition for positivity at the Primary Analysis is statistically compelling evidence of clinically meaningful effects on that primary outcome.

An important design feature when using our OS monitoring guideline is the proposal that the trial be continued post marketing, when feasible and relevant, if the analyses based on $L_{PA}$ deaths lead to regulatory submission and subsequent approval. This enables the trial to provide meaningfully higher statistical power to test the null hypothesis based on ruling out the margin, $δ_{null}$, at the Final Analysis. In turn, to protect the integrity of that assessment, confidentiality of aggregate data by intervention group should be maintained as well as possible after marketing approval (as done when using this design approach in the T2DM clinical setting). A Data Access Plan should prespecify procedures for maintaining confidentiality of data analyzed at the time of the Primary Analysis (Fleming, 2015) where unblinding of aggregate data by intervention group would be done only in a manner to enable regulatory filing. To further enhance trial integrity, these monitoring guidelines should, if possible, be specified prior to any formal interim analysis to avoid post-hoc implementation. Another important design feature would be to ensure study participants continue to be managed according to their original randomization, as has been achieved between the Primary Analysis and Final Analysis in the setting of cardiovascular safety trials conducted in T2DM patients. In particular, allowing control participants to be crossed into the experimental intervention after the Primary Analysis data cut would be problematic (Fleming, Rothmann and Lu, 2009). Consistently avoiding cross-ins may not be possible in all contexts, such as in a setting of high unmet medical where the experimental intervention provides a transformative benefit on the trial's primary 'feels, functions' outcome. In such scenarios the interpretation of OS data collected after the Primary Analysis may be considered compromised, but the proposed OS monitoring guidelines are transparent about the risks associated with decisions made at the Primary Analysis.

The proposed approach in this article is particularly important in the plausible setting where true effects of an intervention on OS are meaningfully inconsistent with the assumption of proportional hazards. In most clinical settings, the standard approach is to estimate the OS HR under a Cox regression proportional hazards model, supported by the absence of prior knowledge that true effects would meaningfully depart from the proportional hazards assumption. Even if that assumption were violated, the HR estimate obtained from fitting a Cox model is still interpretable and relevant, i.e., it is an estimate of the weighted average of the true time-varying HR weighted by where the events occur. Importantly, this estimate is influenced by the censoring distribution. Hence, consider an indolent cancer setting where the true OS HR would be > 1.25-1.50 in the first 24 months post randomization yet < 0.50 thereafter. In that setting, the estimated OS HR readily would exceed the OS HR positivity threshold at the Primary Analysis time, $t_{PA}$, yet would manifestly satisfy that positivity threshold at the Final Analysis time, $t_{FA}$. If such early unfavorable OS data available at the Primary Analysis would influence a regulatory authority to delay approval, continuation of the randomized interventions and capture of longer-term OS data through the Final Analysis time properly would enable subsequent approval of the intervention, meaningfully influenced by the recognition that true overall effects on OS indeed are favorable.

Several important topics are expected expansions of this work, requiring specific deeper developments that are outside the scope of the current paper. Firstly, the proposed OS monitoring guidelines provide prespecified criteria for whether to flag a safety issue. We have, however, not speculated what actions should follow if concerning evidence of a detrimental OS effect is identified at an analysis. For example, consider the scenario where there is a negative OS trend in the overall trial population at the Primary Analysis but there is evidence of a beneficial OS effect in a biologically plausible subgroup identified post-hoc. While post-hoc subgroup analyses should be viewed with caution and usually should be considered to be hypothesis-generating, proper incorporation of exploratory subgroup analyses into OS monitoring guidelines deserves to be a topic of future work.

Secondly, we have framed the monitoring guidelines in this paper in terms of frequentist summaries of the accumulating OS data and positivity thresholds are calibrated to control nominal error probabilities at acceptable values under certain values of the true OS HR. Future work could consider alternative frequentist and Bayesian approaches to formulate guidelines. For example, an anonymous peer reviewer highlighted 'reverse stochastic curtailment' (Tan and Kutner, 1998) as one approach to inform specification of $\ell_{PA}$ without the need for assumptions on the true OS HR. Reverse stochastic curtailment would involve calculating at the Primary Analysis the conditional probability of observing an OS HR estimate less than or equal to the estimate actually seen given the OS HR estimate at the Final Analysis would narrowly exceed the positivity threshold. A small conditional probability could be taken as reassurance that we will be able to 'rule out' $\delta_{null}$ at the level of evidence required at the Final Analysis: $\ell_{PA}$ can be defined as a threshold below which we have sufficient reassurance that the effect on OS does not meet $\delta_{null}$.

Alternatively, decision makers or DMC members may find it informative to see a posterior distribution of the OS HR so that they can identify which regions of the parameter space have high probability given the observed data, even when no relevant trial-external evidence is available, so that Bayesian inference is based on weakly- or non-informative priors. This approach may be a useful supplement to confidence intervals; thus far, we have focused on inspecting whether or not the upper bound of the confidence interval lies below $\delta_{null}$. In addition, the predictive probability of meeting the positivity threshold at the Final Analysis may be a useful measure of risk because it accounts for the uncertainty about the true OS HR that remains at the time of the Primary Analysis. It will often be desirable to integrate multiple sources of information into the evaluation of OS, and Bayesian methods provide a formal framework for doing so. Future work would formulate Bayesian OS monitoring guidelines for cases where relevant and reliable trial-external data can be leveraged to augment the immature OS data from the new trial. Potential areas include using Bayesian approaches to extrapolate the survival curve beyond the range of observation (Guyot et al, 2017), or to incorporate external information on the between-study correlation between treatment effects on PFS and OS in a Bayesian model linking these outcomes.

Randomized clinical trials conducted in chronic indolent cancer settings provide important opportunities for advancing health care in a timely and evidence-based manner by enabling reliable evaluation of benefits and risks of the experimental intervention. It is appropriate that these trials, intended to be confirmatory, e.g., to support either accelerated approval or full approval, often have primary outcomes that are direct measures of how patients 'feel, function or survive', or would be surrogates. The trial should be designed to provide the maximum insights about effects on OS that would be practically achievable. If these OS monitoring guidelines would be implemented at the time of the Primary Analysis and the Final Analysis, then the trial could be conducted in an ethically and scientifically proper manner. In turn, they would provide needed insights for the regulatory approval process and, if the intervention is approved, enable participants to have an informed choice about use of such interventions to improve their health care.

Acknowledgements

We are grateful to acknowledge stimulating discussions with Mouna Akacha and Paul Gallo during the development of this manuscript.

Funding statement

Fleming was partially funded during development of this manuscript by the National Institutes of Health (NIH) grant R37-AI29168. Wittes has received no funding for work on this manuscript.

Disclosure statement

The authors report there are no competing interests to declare. Hampson, Bharani-Dharan, Bretz, Chakravartty, Coroller, Koukouli, Yateman and Zuber are employees of Novartis and own stocks in this company. Wittes has consulting agreements with Amgen, Anthos, BMS, Astra-Zeneca, Glycomimetics, Cell Prothera, NED, Novartis, all of which are developing products in oncology. Fleming has a research consultancy agreement with Novartis. He also is a consultant to the National Institutes of Health and the pharmaceutical and biotech industry in the design, monitoring, and analysis of clinical trials. He holds no stock in pharmaceutical or biotech companies.

# Appendix A: Derivation of expressions for positivity thresholds for OS monitoring guidelines

For simplicity, consider a trial that has two looks at the OS data, at the Primary and Final Analyses. Suppose patients are randomized to the experimental intervention and control in a k to 1 ratio. Furthermore, let $\hat{\theta}_{PA}$ and $\hat{\theta}_{FA}$ denote estimators of the OS log-HR at the Primary and Final Analysis, respectively. Asymptotically, $\hat{\theta}_{PA} \sim N(\theta, \mathcal{J}_{PA}^{-1})$ and $\hat{\theta}_{FA} \sim N(\theta, \mathcal{J}_{FA}^{-1})$, where the Fisher Information levels $\mathcal{J}_{PA} = kL_{PA}/(k+1)^2$ and $\mathcal{J}_{FA} = kL_{FA}/(k+1)^2$ (Gsponer et al, 2014), at least approximately assuming the OS HR is close to 1. For more extreme OS HRs outside of 0.5 to 2.0, adjustment factors should be applied to these expressions for the Fisher Information levels; see Equation (5) of Al-Khalidi et al (2011) for the adjustment factor when k=1. In what follows, we derive expressions for the positivity thresholds shown in Section 3.2 for the OS HR point estimates for the case that k=1.

*Final Analysis:*

Given $L_{PA}$, the positivity threshold is fixed by specification of $\gamma_{FA}$. Let $\ell_{FA}$ denote the positivity threshold on the log-HR scale. Then $\ell_{FA}$ must satisfy $\Pr\{\hat{\theta}_{FA} < \ell_{FA}; \theta = \log(\delta_{null})\} = \gamma_{FA}$. Straightforward algebra shows that $\ell_{FA} = \log(\delta_{null}) + (k+1)\Phi^{-1}(\gamma_{FA})/\sqrt{kL_{FA}}$, where $\Phi$ is the cumulative distribution function of a standard normal random variate. Taking exponentials, we obtain $\exp(\ell_{FA}) = \delta_{null}\exp\left\{\frac{(k+1)}{\sqrt{k}}\frac{\Phi^{-1}(\gamma_{FA})}{\sqrt{L_{FA}}}\right\}$.

*Primary Analysis*

Suppose we phrase the positivity criterion at the Primary Analysis as 'the upper bound of the (1 − 2$\gamma_{PA}$)100% 2-sided CI for the OS log-HR should be less than $\log(\delta_{null})$'. In order to achieve the target marginal power, we must choose $\gamma_{PA}$ such that $\Pr\{\hat{\theta}_{PA} + \Phi^{-1}(1-\gamma_{PA})(k+1)/\sqrt{kL_{PA}} <$ $\log(\delta_{\text{null}})$; $\theta = \log(\delta_{\text{alt}})\} = 1 - \beta_{PA}$. Rearranging for γ_PA, we obtain $\gamma_{PA} = 1 - \Phi\left(\frac{\sqrt{kL_{PA}}\log(\delta_{\text{null}}/\delta_{\text{alt}})}{(k+1)} - \Phi^{-1}(1 - \beta_{PA})\right)$. Denoting the threshold for positivity for the OS log-HR point estimate by $\ell_{PA}$, this threshold must satisfy:

$$\ell_{PA} + \Phi^{-1}(1 - \gamma_{PA})(k+1)/\sqrt{kL_{PA}} = \log(\delta_{\text{null}})$$

Rearranging, the positivity threshold on the OS HR scale is $\exp(\ell_{PA}) = \delta_{\text{null}}\exp\left\{\frac{(k+1)}{\sqrt{kL_{PA}}}\Phi^{-1}(\gamma_{PA})\right\}$.

Figure 1: Schematic diagram illustrating the timing of the Primary Analysis of the trial's primary outcome and the Final Analysis for OS, along with the number of OS events at each of these two key calendar times.

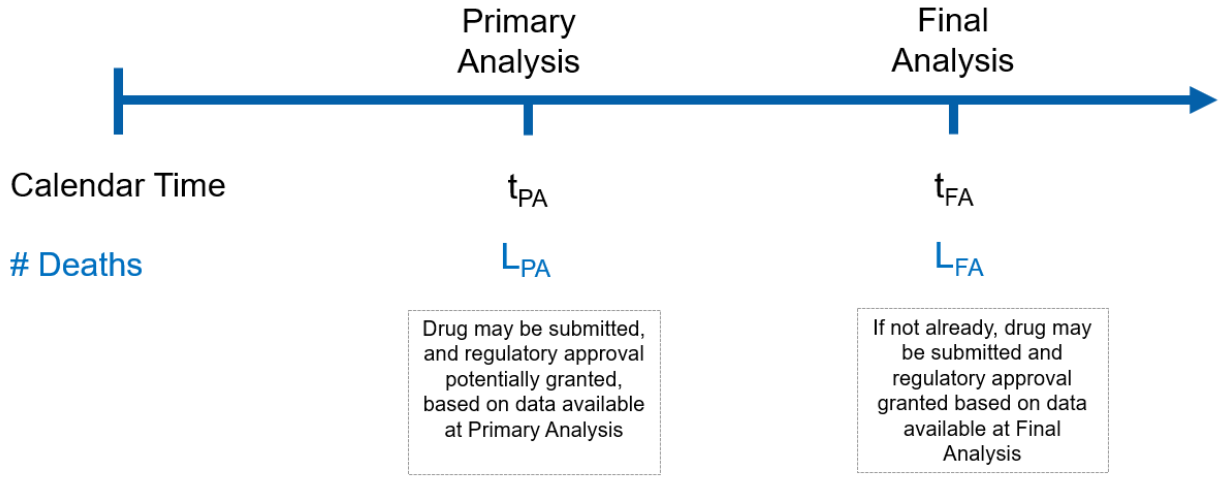

Table 1: OS Detriments Observed Across the PI3K Inhibitor Class. Source: FDA Introductory Comments presented at April 21st 2022 ODAC on PI3K Inhibitors (FDA ODAC, 2022)

| Study | Population & Treatment | Deaths | | Hazard Ratio (95% CI) |
|---|---|---|---|---|
| | | PI3Ki arm | Control arm | |
| 312-0123 | Untreated CLL<br>*Bendamustine & rituximab ± idelalisib* | 8% (12/157) | 3% (4/154) | 3.34 (1.08, 10.39) |
| 313-0124 | Previously treated indolent NHL[2]<br>Rituximab ± idelalisib | 5% (10/191) | 1% (1/95) | 4.74 (0.6, 37.12) |
| 313-0125 | Previously treated indolent NHL<br>*Bendamustine & rituximab ± idelalisib* | 8% (27/320) | 6% (9/155) | 1.51 (0.71, 3.23) |
| DUO | Previously treated CLL/SLL[1]<br>*Duvelisib vs ofatumumab* | 50% (80/160) | 44% (70/159) | 1.09 (0.79, 1.51) |
| CHRONOS-3 | Previously treated indolent NHL<br>*Rituximab ± copanlisib* | 18% (56/307) | 21% (32/151) | 0.87 (0.57, 1.35) |
| UNITY-CLL | Untreated and previously treated CLL<br>*Umbralisib + ublituximab vs GC[3]* | * | * | 1.23 |

[1]CLL/SLL, chronic lymphocytic leukemia/small lymphocytic leukemia
[2]NHL, non-Hodgkin lymphoma
[3]GC, obinutuzumab and chlorambucil
* Not publicly available

Table 2: Glossary of notation and terminology. Calendar times are measured relative to study initiation.

| Term | Interpretation |
|---|---|
| Indolent cancer | A cancer characterized by slow growth and progression and low mortality rate. |
| $δ_{null}$ | Smallest unacceptable detrimental OS HR. |
| $δ_{alt}$ | Plausible OS HR consistent with (perhaps incremental) benefit that is reasonably expected from the experimental intervention. |
| Primary Analysis | The trial's decisive analysis of the primary outcome at which a definitive conclusion on the primary null hypothesis is reached. |
| $t_{PA}$ | Time (months) of the Primary Analysis. |
| $L_{PA}$ | Number of deaths we expect by time $t_{PA}$. This usually is not a design feature and is based on assumptions on OS distributions. |
| Final Analysis | The final analysis of OS. This analysis may occur post-marketing. |
| $t_{FA}$ | Time (months) of the Final Analysis. |
| $L_{FA}$ | Number of deaths we expect by time $t_{FA}$. |
| Positivity threshold | At an analysis, this is the value below which the observed OS HR (test/control) must lie to provide sufficient reassurance that the novel intervention does not have an unacceptable detriment. |
| $\ell_{PA}$ | Value below which the observed OS HR must lie at the Primary Analysis to provide sufficient reassurance that the effect on OS does not reach $δ_{null}$. |
| $\ell_{FA}$ | Value below which the observed OS HR must lie at the Final Analysis to provide sufficient reassurance that the effect on OS does not reach $δ_{null}$. |
| False positive error | Made when the true OS HR is $δ_{null}$ but the observed OS HR lies below the positivity threshold. The practical implication of making this error is that we are incorrectly reassured that a harmful novel intervention is adequately safe. |
| $Υ_{PA}$ | False positive error rate at the Primary Analysis. |
| $Υ_{FA}$ | False positive error rate at the Final Analysis. |
| False negative error | Made when the true OS HR is $δ_{alt}$ but the observed OS HR exceeds the positivity threshold. The practical implication of making this error is that we incorrectly flag concerning evidence of a detrimental effect when in fact the intervention has incremental benefit. |
| $\beta_{PA}$ | False negative error rate at the Primary Analysis. |
| $\beta_{FA}$ | False negative error rate at the Final Analysis. |

Table 3: OS monitoring guideline retrospectively applied to Motivating Example 1 with $\delta_{null}$ = 1.3, $\delta_{alt}$ = 0.80, $\gamma_{FA}$ = 0.025 and $\beta_{PA}$ = 0.10.

| # deaths at Analysis | OS HR threshold for positivity | One-sided false positive error rate | Level of 2-sided CI needed to rule out $\delta_{null}$ | Probability of meeting positivity threshold under $\delta_{alt}$ | POLARIX OS results ($\widehat{HR}$) |
|---|---|---|---|---|---|
| 60  | 1.114 | 0.275 | 45% | 0.900 |      |
| 89  | 1.050 | 0.157 | 69% | 0.900 |      |
| 110 | 1.021 | 0.103 | 79% | 0.900 | 0.94 |
| 131 | 1.001 | 0.067 | 87% | 0.900 | 0.94 |
| 178 | 0.969 | 0.025 | 95% | 0.900 | ---  |

Table 4: OS monitoring guideline with $\delta_{null}$ = 1.3, $\delta_{alt}$ = 0.70, $\gamma_{FA}$ = 0.10 and $\beta_{PA}$ = 0.10.

| # deaths at Analysis | OS HR threshold for positivity | One-sided false positive error rate | Level of 2-sided CI needed to rule out $\delta_{null}$ | Probability of meeting positivity threshold under $\delta_{alt}$ | Probability of meeting positivity threshold under HR = 0.95 |
|---|---|---|---|---|---|
| 28 | 1.136 | 0.361 | 28% | 0.900 | 0.682 |
| 42 | 1.040 | 0.234 | 53% | 0.900 | 0.615 |
| 70 | 0.957 | 0.100 | 80% | 0.905 | 0.512 |

Table 5: OS monitoring guideline with $\delta_{null}$ = 1.3, $\delta_{alt}$ = 0.95, $\gamma_{FA}$ = 0.2 and $\beta_{PA}$ = 0.25.

| # deaths at Analysis | OS HR threshold for positivity | One-sided false positive error rate | Level of 2-sided CI needed to rule out $\delta_{null}$ | Probability of meeting positivity threshold under $\delta_{alt}$ | Probability of meeting positivity threshold under HR = 1 |
|---|---|---|---|---|---|
| 28 | 1.226 | 0.438 | 12% | 0.750 | 0.705 |
| 42 | 1.170 | 0.366 | 27% | 0.750 | 0.694 |
| 70 | 1.063 | 0.200 | 60% | 0.681 | 0.601 |

Table 6: OS monitoring guideline applied to Motivating Example 2 with $\delta_{null}$ = 1.333, $\delta_{alt}$ = 0.7, $\gamma_{FA}$ = 0.20 and $\beta_{PA}$ = 0.1.

| Follow-up (months) | # deaths at Analysis | OS HR threshold for positivity | One-sided false positive error rate | Level of 2-sided CI needed to rule out $\delta_{null}$ | Probability of meeting positivity threshold under $\delta_{alt}$ | Probability of meeting positivity threshold under HR = 0.95 |
|---|---|---|---|---|---|---|
| 55 ($t_{PA}$) | 22 | 1.209 | 0.409 | 18% | 0.900 | 0.714 |
| 87 ($t_{FA}$) | 34 | 0.999 | 0.200 | 60% | 0.850 | 0.558 |